\documentclass[12pt,a4]{article} 
\newcommand{\vs}{\vspace{-0.25cm}}
\usepackage{amsmath,graphicx}
\begin{document} 
\begin{center}
  {\Large{\bf Solving the matrix exponential function for the groups SU(3), SU(4) and Sp(2)} }  

\medskip

 Norbert Kaiser\\
\medskip
{\small Physik-Department T39, Technische Universit\"{a}t M\"{u}nchen,
   D-85748 Garching, Germany\\

\smallskip

{\it email: nkaiser@ph.tum.de}}
\end{center}
\medskip
\begin{abstract}
The well known analytical formula for $SU(2)$ matrices $U = \exp(i \vec \tau
\!\cdot\! \vec \varphi\,) = \cos|\vec \varphi\,|  + i\vec \tau \!\cdot\! \hat\varphi \, \sin|\vec
\varphi\,|$\\ is extended to the $SU(3)$ group with eight real parameters. The resulting analytical formula involves the sum over three real roots of a cubic equation, corresponding to the so-called irreducible case, where  one has to employ the trisection of an angle. When going to the special unitary group $SU(4)$ with 15 real prameters, the analytical formula involves the sum over four real roots of a quartic equation. The associated cubic resolvent equation with three positive roots belongs again to the irreducible case. Furthermore, by imposing the pertinent condition on $SU(4)$ matrices one can also treat the symplectic group $Sp(2)$ with ten real parameters. Since there the roots occur as two pairs of opposite sign, this simplifies the analytical formula for $Sp(2)$ matrices considerably.    
An outlook to the situation with analytical formulas for $SU(5)$, $SU(6)$ and $Sp(3)$ is also given.\end{abstract}

\section{Introduction and summary}
In chiral effective field theories for low-energy quantum chromodynamics \cite{ulfbook} one typically works with special unitary matrices $U$ as the field variable, since this allows for a convenient implementation of (chiral and other) symmetry transformations. The excitations on top of the spontaneously broken groundstate, represented by the unit matrix $\bf{1}$, are pseudoscalar Goldstone bosons (pions, kaons and the $\eta(548)$-meson). In the two-flavor case the  special unitary  $2\times 2$ matrices forming the compact Lie group $SU(2)$ are often given in the exponential form 
\begin{equation} U = \exp(i \vec \tau\!\cdot \!\vec \varphi\,) = \cos|\vec \varphi\,|{\bf 1} + i\vec \tau \!\cdot\! \hat \varphi \sin|\vec\varphi\,|\,,\end{equation} 
with $\vec \tau=(\tau_1,\tau_2,\tau_3)$ the Pauli matrices and $|\vec \varphi\,|$ the magnitude of a three-component real vector $\vec \varphi=(\varphi_1,\varphi_2,\varphi_3)$ related  e.g. to the pion-fields. Since the manifold of $SU(2)$ is identified with a three-sphere $S^3$, one can provide alternative algebraic or rational parametrizations
\begin{equation} U = \pm \sqrt{1-\vec \pi^{\,2}}\,{\bf 1} + i\vec \tau \!\cdot\! \vec \pi\,, \quad|\vec \pi\,|\leq 1\,,  \qquad U = {(1-\vec \phi^{\,2}/4){\bf 1}+ i\vec \tau \!\cdot\! \vec \phi\over 1+\vec \phi^{\,2}/4} \,,\end{equation} 
which are advantageous in calculations or for certain applications. In the three-flavor case, where the matrices $U$ belong to the eight-dimensional compact Lie group $SU(3)$, no such alternatives to parametrize the manifold are known and one has to stay with the exponential form $U = \exp(i \vec \lambda\!\cdot \! \vec v\,)$ in terms of eight Gell-Mann matrices $\vec \lambda = (\lambda_1,\dots, \lambda_8)$ and eight real parameters $\vec v=(v_1,\dots , v_8)$. The aim of the present work is solve the corresponding matrix exponential function. A key ingredient to limit the number of matrix powers to few independent ones is the Cayley-Hamilton relation, which states that any matrix $\Sigma$ gets nullified when inserted into its characteristic polynomial $P(z)={\rm det}(z \mathbf{1}-\Sigma)$. The coefficients \cite{koecher} of the latter are given by the traces of increasing matrix powers of $\Sigma$ and ultimately the determinant of $\Sigma$. When carrying out this procedure for $SU(3)$, one encounters the problem of determining the roots of a cubic polynomial in the so-called irreducible case. It corresponds to the situation when all three roots are real and the Cardano formula exhibits under the cube-root a square-root with a negative radicand. Then the problem gets effectively solved through a trigonometric ansatz and the trisection of an angle. When continuing the solution of the matrix exponential function to $SU(4)$, the four real roots of a quartic polynomial are determined with the help of a cubic resolvent polynomial that also belongs to the irreducible case. As an interesting byproduct of this analysis one obtains the shape of the allowed region for certain invariants $\eta$ and $\zeta$.  Moreover, be imposing the condition related to a quaternionic structure one can treat as a subgroup of $SU(4)$ the symplectic group $Sp(2)$ with ten real parameters. The situation with more elaborate quasi-analytical formulas for $SU(5)$, $SU(6)$ and $Sp(3)$ matrices in the exponential parametrization is discussed in perspective.        
 \section{Special unitary group SU(3)}
One starts with the usual exponential representation of an $SU(3)$ matrix 
\begin{equation}U = \exp(i \vec \lambda\!\cdot \! \vec v\,)\,,\end{equation}  
in terms of the eight Gell-Mann matrices\footnote{The two diagonal Gell-Mann matrices are $\lambda_3 ={\rm diag}(1,-1,0)$ and  $\lambda_8 = {\rm diag}(1,1,-2)/\sqrt{3}$, while the other six have two non-zero entries either $1,1$ or $-i,i$ placed symmetrically at positions above and below the diagonal.} $\vec \lambda = (\lambda_1,\dots, \lambda_8)$ normalized to tr$( \lambda_a \lambda_b)  =2 \delta_{ab}$ and an eight-component real parameter vector $\vec v = (v_1, \dots ,v_8)$. The aim is to give an analytical expression for $U$ that involves the $3\!\times \!3$ unit matrix $\mathbf{1}$ and the rescaled matrix $\Sigma = \vec \lambda \cdot \hat v$ (where  $\hat v = \vec v/|\vec v\,|$), each multiplied with coefficient functions that depend on the magnitude $|\vec v\,|=\sqrt{v_1^2+\dots+v_8^2}$ and another invariant. Considering the traces tr$\,\Sigma=0$ and  tr$\,\Sigma^2=2$ one finds as a pertinent invariant the (real-valued) determinant $\eta = {\rm det}\,\Sigma$. According to the Cayley-Hamilton relation the traceless hermitian matrix $\Sigma$ (a Lie algebra element) gets nullified when inserted into its characteristic polynomial, which lead to the cubic relation
\begin{equation} \Sigma^3 = \Sigma + \eta\, \mathbf{1}\,.\end{equation} 
Consequently, any power of $\Sigma$ can be written as a linear combination of  $\mathbf{1}, \Sigma$, and  $\Sigma^2$. Starting at order $n$ with $\Sigma^n =\alpha_n \, \mathbf{1}+\beta_n\,  \Sigma +\gamma_n\,  \Sigma^2 $ and multiplying with $\Sigma$ one obtains via the relation in eq.(4) the expansion coefficients at order $n+1$. The resulting linear recursion relation reads in vector notation:
\begin{equation} \left(\!\begin{array} {c}\alpha_{n+1} \\ \beta_{n+1} \\ \gamma_{n+1} \cr\end{array}\!\right) = M_3  \left(\!\begin{array} {c}\alpha_n \\ \beta_n \\ \gamma_n \cr\end{array}\!\right), \quad \text{ with the matrix} \quad M_3 =  \left( \begin{array} {ccc} 0 & 0& \eta 
\\ 1& 0& 1\\ 0&1& 0 \cr \end{array}\!\right), 
\end{equation} 
and the initial values $\alpha_0=1, \beta_0=0, \gamma_0=0$.  
Through diagonalization of $M_3$ the exponential series $\exp( i|\vec v\,|M_3) = \sum_{n=0}^\infty  ( i|\vec v\,|M_3)^n/n!$ can be solved\footnote{When using Mathematica, the routine MatrixExp[ , ] gives the result directly in terms of RootSum[ , ].} and by dotting with $(1,0,0)^t$ from the right and $( \mathbf{1}, \Sigma,\Sigma^2)$ from the left, one ends up with the following analytical formula
for an $SU(3)$ matrix
\begin{equation}U = \exp(i \vec \lambda\!\cdot \! \vec v\,) = \sum_{j=1}^3 {\exp(i z_j |\vec v\,|) \over 3z_j^2-1} \Big\{ (z_j^2-1)  \mathbf{1}+ z_j \, \vec \lambda\!\cdot \! \hat v+ (\vec \lambda\!\cdot \! \hat v)^2 \Big\}\,,
 \end{equation}
where the $\eta$-dependent quantities $z_1,z_2,z_3$ subject to the zero-sum constraint $z_1+z_2+z_3=0$  are the three real roots of the cubic equation
 \begin{equation} P_3(z) = z^3-z-\eta = 0\,. \end{equation}
 
 \begin{figure}\centering
\includegraphics[width=0.45\textwidth]{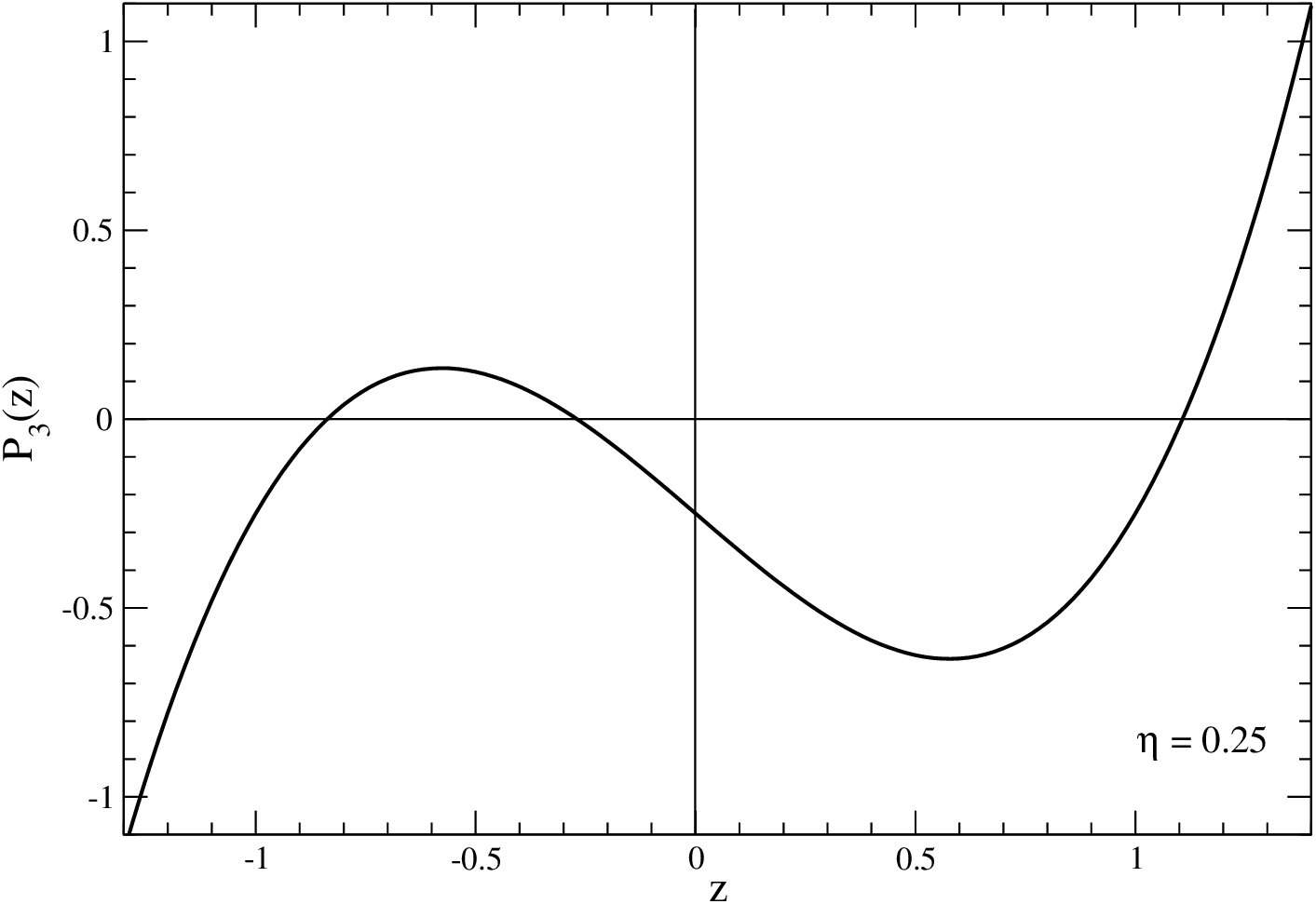}
\caption{Generic behavior of the cubic polynomial $P_3(z) = z^3-z -\eta$.} 
\end{figure}
\noindent Note that at the same time $z_1,z_2,z_3$ are the eigenvalues of the traceless hermitian matrix $\Sigma = \vec \lambda\!\cdot \! \hat v$. One also observes that the trace of $U$ is given by a much simpler formula: tr$\,U =  \sum_{j=1}^3 \exp(i z_j |\vec v\,|)$. Another interesting feature is that the denominator in eq.(6) is the derivative $P_3'(z) = 3z^2-1$ evaluated at the root $z_j$. Before turning to the solution of the cubic equation, one should analyze the generic behavior of the polynomial $P_3(z)= z^3-z-\eta$ as it is sketched in Fig.\,1.  When having three zero-crossings, the local maximum at $z= -1/\sqrt{3}$ must be positive, $2\sqrt{3}/9-\eta \geq0$, and the local minimum  at $z= 1/\sqrt{3}$ must be negative, $-2\sqrt{3}/9-\eta \leq 0$. Both conditions together fix the range of the determinant $\eta$ to the interval  
 \begin{equation} -{2\sqrt{3} \over 9}\leq \eta \leq {2\sqrt{3} \over 9}\,. \end{equation}
In the irreducible case at present, the cubic equation $z^3-z-\eta=0$ is treated by the substitution $z ={2 \over \sqrt{3}} \cos\psi$ which leads (via the relation $\cos3\psi = 4\cos^3\psi-3\cos\psi$) to the auxiliary equation 
  \begin{equation} \cos3\psi  = {3\sqrt{3} \eta\over 2}\,, \qquad \psi  = {1\over 3} \arccos{3\sqrt{3} \eta\over 2}\,, \end{equation}
that  provides an angle $ \psi$ in the range $0\leq \psi \leq \pi/3$. It is an important solvability criterion that the argument in eq.(9) has indeed magnitude less equal to 1 as a consequence of the determinantal range derived before. The three real roots entering the analytical formula in eq.(6) are given by
  \begin{equation} z_1 = {2\over \sqrt{3}} \cos\psi >0\,, \quad z_2  = -\sin\psi -{\cos\psi \over \sqrt{3}}<0\,,\quad \quad z_3  = \sin\psi -{\cos\psi \over \sqrt{3}}\,, \end{equation} 
 where in this (arbitrary) ordering the sign of $z_3$ is yet undetermined. The validity of the quasi-analytical formula has been checked numerically in many cases. A disadvantage for applications is its rather inexplicit dependence on $\eta$ which cannot be expressed in terms of real-valued algebraic functions, but requires a detour via the real roots of a cubic equation in the irreducible case.   For comparison in the case of $SU(2)$ the iteration matrix $M_2$ is the Pauli matrix $\tau_1$ and $\exp( i |\vec\varphi\,|\tau_1)(1,0)^t = (\cos |\vec\varphi\,|,  i\sin|\vec\varphi\,|)^t$.

\section{Special unitary group SU(4)}
One starts again with the exponential representation of an $SU(4)$ matrix 
\begin{equation}U = \exp(i \vec \lambda\!\cdot \! \vec v\,)\,,\end{equation}  
in terms of the 15  traceless hermitian generators\footnote{The three diagonal generators are $\lambda_3 ={\rm diag}(1,-1,0,0),\, \lambda_8 = {\rm diag}(1,1,-2,0)/\sqrt{3},$ and $ \lambda_{15} = {\rm diag}(1,1,1,-3)/\sqrt{6}$. The remaining 12 generators have two non-zero entries either $1,1$ or $-i,i$ placed symmetrically at positions above and below the diagonal.}  $\vec \lambda = (\lambda_1,\dots, \lambda_{15})$ normalized to tr$( \lambda_a \lambda_b)  =2 \delta_{ab}$ and a 15-component real parameter vector $\vec v = (v_1, \dots ,v_{15})$. Again, one works with the $4\!\times\! 4$ unit matrix $\mathbf{1}$ and the rescaled matrix $\Sigma = \vec \lambda\!\cdot \! \hat v$ that get multiplied by functions depending  on the magnitude $|\vec v\,|=\sqrt{v_1^2+\dots+v_{15}^2}$ and further invariants. Besides the constant traces tr$\,\Sigma=0$ and  tr$\,\Sigma^2=2$ one finds now as the two pertinent (real-valued) invariants the determinant and the trace of the cube
 \begin{equation}\eta = {\rm det}\,\Sigma\,, \qquad \zeta = {1\over 3} {\rm tr}\,\Sigma^3\,,\end{equation}
 where the factor $1/3$ is included for convenience. In present case the Cayley-Hamilton relation sets up an equation for the fourth power of $\Sigma$ of the form
\begin{equation} \Sigma^4 = \Sigma^2 +\zeta \,\Sigma  - \eta\, \mathbf{1}\,,\end{equation} 
which allows to write any higher power of $\Sigma$ as a linear combination of  $\mathbf{1}, \Sigma, \Sigma^2$, and $\Sigma^3$. Starting at order $n$ with $\Sigma^n =\alpha_n \, \mathbf{1}+\beta_n\,  \Sigma +\gamma_n\,  \Sigma^2 +\delta_n\,  \Sigma^3$ and multiplying with $\Sigma$ one obtains via the relation in eq.(13) the expansion coefficients at order $n+1$. The resulting linear recursion relation reads in vector notation:
\begin{equation} \left(\!\begin{array} {c}\alpha_{n+1} \\ \beta_{n+1} \\ \gamma_{n+1} \\ \delta_{n+1} \cr\end{array}\!\right) = M_4  \left(\!\begin{array} {c}\alpha_n \\ \beta_n \\ \gamma_n \\ \delta_n \cr\end{array}\!\right), \quad \text{ with the  matrix} \quad M_4 =  \left( \begin{array} {cccc} 0& 0 & 0& -\eta \\ 1& 0& 0& \zeta \\ 0&1& 0&1\\ 0 & 0 &1&0 \cr \end{array}\!\right), 
\end{equation} 
and the initial values $\alpha_0=1, \beta_0=0, \gamma_0=0, \delta_0=0$.  
Through diagonalization of $M_4$ (or an application of MatrixExp[ , ]) the exponential series $\exp( i|\vec v\,|M_4) = \sum_{n=0}^\infty  ( i|\vec v\,|M_4)^n/n!$ can again be solved  and by dotting with $(1,0,0,0)^t$ from the right and $( \mathbf{1}, \Sigma,\Sigma^2, \Sigma^3)$ from the left, one ends up with the following quasi-analytical formula for an $SU(4)$ matrix
\begin{equation}U = \exp(i \vec \lambda\!\cdot \! \vec v\,) = \sum_{j=1}^4 {\exp(i z_j |\vec v\,|) \over 4z_j^3-2z_j-\zeta} \Big\{ (z_j^3-z_j-\zeta)\mathbf{1}+ (z_j^2-1) \vec \lambda\!\cdot \! \hat v+ z_j  (\vec \lambda\!\cdot \! \hat v)^2+ (\vec \lambda\!\cdot \! \hat v)^3 \Big\}\,,
 \end{equation}
 where the $(\eta,\zeta)$-dependent quantities $z_1,z_2,z_3,z_4$ subject to the zero-sum constraint $z_1+z_2+z_3+z_4=0$  are now the four real roots of the quartic equation
 \begin{equation} P_4(z) = z^4-z^2- \zeta\,z +\eta = 0\,. \end{equation}
At the same time $z_1,z_2,z_3,z_4$ are the eigenvalues of the traceless hermitian $4\!\times \!4$ matrix $\Sigma = \vec \lambda\!\cdot \! \hat v$ and \\ one gets again a simpler formula for the trace: tr$\,U =  \sum_{j=1}^4 \exp(i z_j |\vec v\,|)$. The denominator in eq.(15) stems from the derivative $P_4'(z) = 4z^3-2z-\zeta$. 

Borrowing results from advanced algebra \cite{algebra}, the determination of the four roots $z_j$ of the quartic equation $P_4(z)=0$ proceeds via three auxiliary quantities $\theta_1,\theta_2,\theta_3$ in the following way

\begin{figure}[ht]\centering
\includegraphics[width=0.45\textwidth]{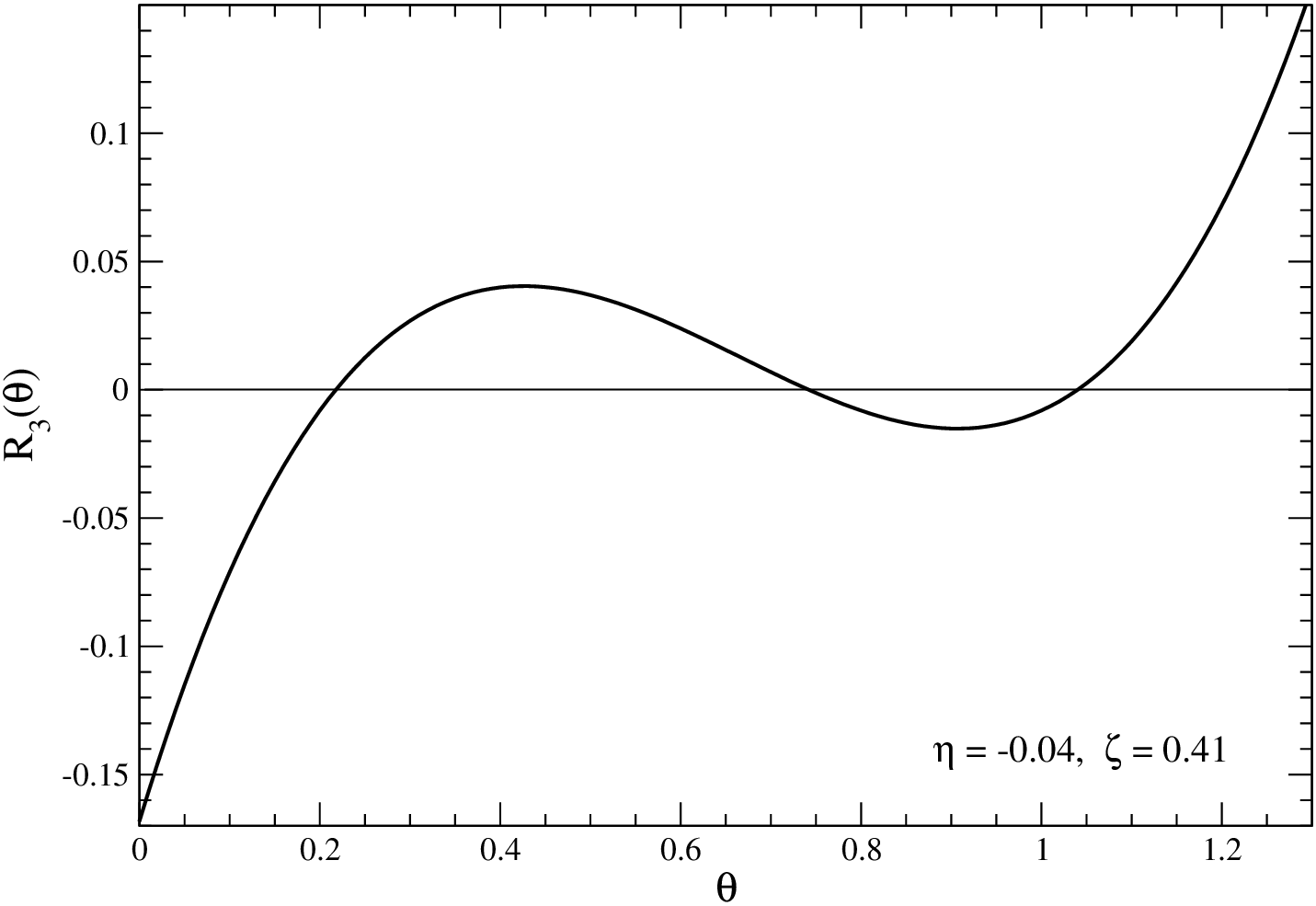}
\caption{Generic behavior of the cubic resolvent polynomial $R_3(\theta) = \theta^3-2\theta^2+(1-4\eta)\theta-\zeta^2$.} 
\end{figure}

\begin{equation} z_1 = {1\over 2}\Big(\sqrt{\theta_1}+ \sqrt{\theta_2}+ \sqrt{\theta_3} \Big)\,, \qquad z_2 = {1\over 2}\Big( \sqrt{\theta_1}- \sqrt{\theta_2}- \sqrt{\theta_3}\Big)\,,  
\end{equation}
\begin{equation} z_3 = {1\over 2}\Big(-\sqrt{\theta_1}+ \sqrt{\theta_2}- \sqrt{\theta_3} \Big)\,, \qquad z_4 = {1\over 2}\Big(-\sqrt{\theta_1}- \sqrt{\theta_2}+ \sqrt{\theta_3}\Big)\,,\end{equation}
where it has to be noted that there is only a twofold sign ambiguity in taking square roots, since the sign of the product is fixed by the condition $\sqrt{\theta_1} \sqrt{\theta_2} \sqrt{\theta_3} = \zeta$. The four choices of signs $+\,+, +\,-, -\,+, -\,-$ correspond merely to a relabeling of the four roots. The $\theta$-values derive from the roots through the inverse relations 
\begin{equation} \theta_1 = -(z_1+z_2)(z_3+z_4)\,, \qquad  \theta_2 = -(z_1+z_3)(z_2+z_4)\,, \qquad  \theta_3 = -(z_1+z_4)(z_2+z_3)\,,  \end{equation} 
and these are all positive, since each is a square in view of the zero-sum  $z_1+z_2+z_3+z_4=0$. As a matter of fact \cite{algebra} the three $\theta$-values are the roots of the cubic resolvent equation \cite{algebra}
\begin{equation} R_3(\theta) = \theta^3 -2 \theta^2+(1-4\eta)\theta - \zeta^2 =0\,. \end{equation} 
The generic behavior of $ R_3(\theta)$ which is shown in Fig.\,2 implies several restrictions on the invariants $\eta$ and $\zeta$. In the presence of three zero-crossing on the positive $\theta$-axis and $R_3(\theta<0)<-\zeta^2$,  the local maximum and local minimum must lie in between at positions $\theta_\text{max,min} = {1\over 3} (2\mp \sqrt{1+12 \eta}) \geq 0$. This implies first $\eta\geq -1/12$ and secondly $\eta \leq 1/4$, leading to the (narrow) range $-1/12\leq \eta\leq1/4$ for the determinant. The conditions $R_3(\theta_\text{max}) \geq 0$ and $R_3(\theta_\text{min}) \leq 0$ multiplied with each other lead after some manipulation to the inequality
 \begin{equation} \Big({27\over 2}\zeta^2+36 \eta-1\Big)^2 \leq (1+12\eta)^3 \leq 64\,, \end{equation} 
The resulting allowed range for  the invariants $\eta$ and $\zeta$ is the bounded region shown in Fig.\,3 from which one deduces also the extremal values $\zeta_\pm = \pm 2 \sqrt{6}/9$.
\begin{figure}[ht]\centering
\includegraphics[width=0.45\textwidth]{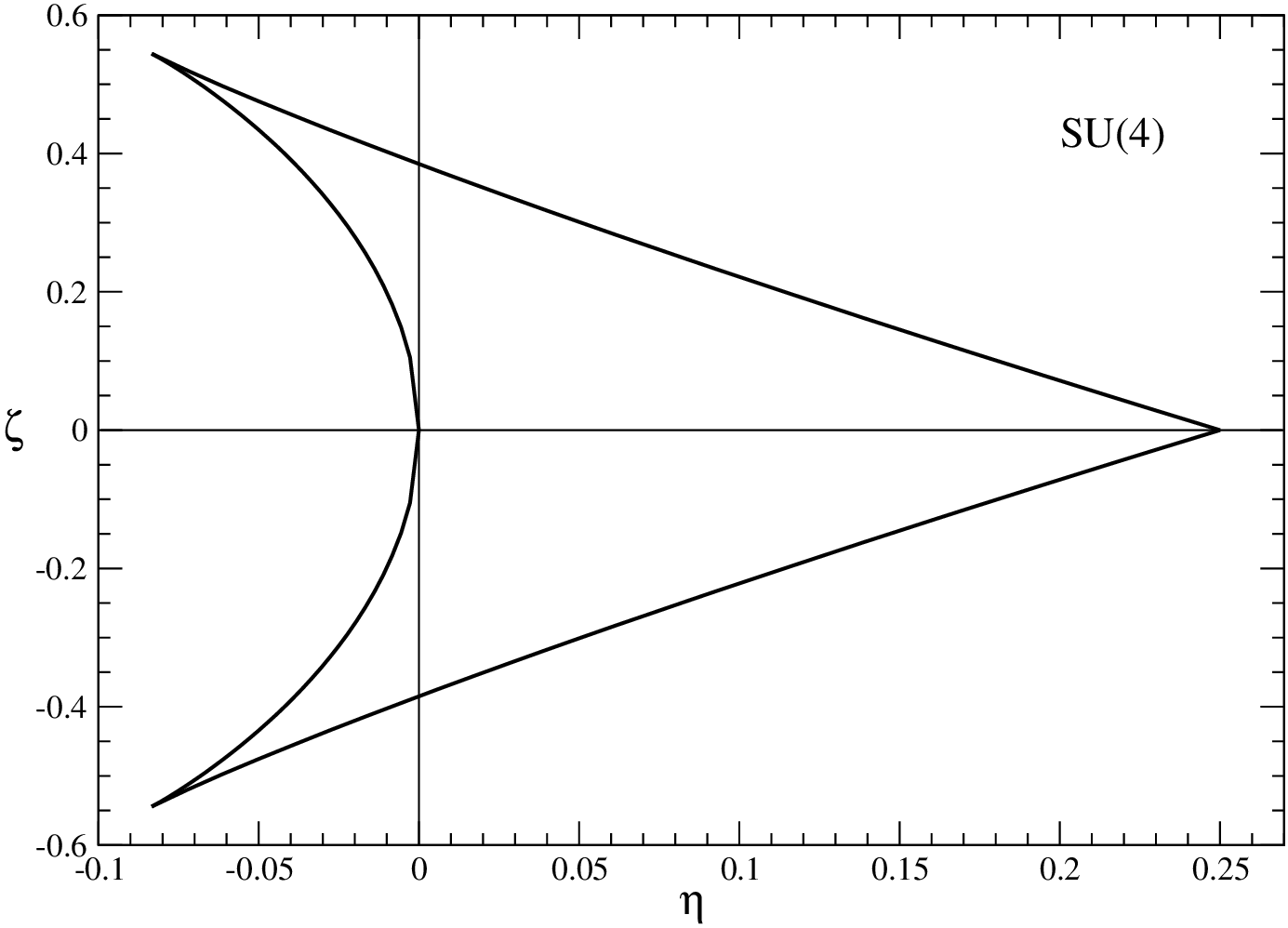}
\caption{The allowed values of the invariants  $\eta$ and $\zeta$  lie inside the bounded region.} 
\end{figure}

The cubic resolvent equation in eq.(20) belongs again to the irreducible case such that its solutions are conveniently obtained via the substitution $\theta = {2\over 3}(1+\sqrt{1+12\eta} \cos\psi)$. The auxiliary angle $\psi$ lying within the interval $[0,\pi/3]$ is determined from the equation
\begin{equation} \cos3\psi = {{27\over 2}\zeta^2+36 \eta-1\over (1+12\eta)^{3/2}}\,, \qquad \psi = {1\over 3} \arccos{{27\over 2}\zeta^2+36 \eta-1\over (1+12\eta)^{3/2}}\,, \end{equation} 
where solvability is guaranteed by the inequality derived previously  in eq.(21). The three positive $\theta$-values read (up to permutation of the indices)
\begin{equation}\theta_1 = {2\over 3}(1+\sqrt{1+12\eta} \cos\psi)\,, \quad \theta_2 = {2\over 3}\Big[1-\sqrt{1+12\eta} \sin\Big(\psi+{\pi \over 6}\Big)\Big] \,, \quad \theta_3 = {2\over 3}\Big[1+\sqrt{1+12\eta} \sin\Big(\psi-{\pi \over 6}\Big)\Big] \,, \end{equation}
and after taking square roots and forming appropriate sums and differences as prescribed in eqs.(17,18) one obtains the four real roots $z_j$ entering the quasi-analytical formula eq.(15) for an $SU(4)$ matrix.

One can continue the procedure to $SU(5)$  with 24  traceless hermitian generators\footnote{The four diagonal generators are $\lambda_3 ={\rm diag}(1,-1,0,0,0),\, \lambda_8 = {\rm diag}(1,1,-2,0,0)/\sqrt{3},\, \lambda_{15} = {\rm diag}(1,1,1,-3,0)/\sqrt{6},$ and $\lambda_{24} = {\rm diag}(1,1,1,1,-4)/\sqrt{10}$. The remaining 20 generators have two non-zero entries  either $1,1$ or $-i,i$ placed symmetrically at positions above and below the diagonal.} $\vec \lambda = (\lambda_1,\dots, \lambda_{24})$ normalized to tr$( \lambda_a \lambda_b)  =2 \delta_{ab}$ and a 24-component real parameter vector $\vec v = (v_1, \dots ,v_{24})$. In terms of the invariants $\eta={\rm det}\,\Sigma, ~ \zeta={\rm tr}\,\Sigma^3/3$ and a properly chosen new one, $\xi=  {\rm tr}\,\Sigma^4/4-1/2$, the Cayley-Hamilton relation for the fifth power of $\Sigma = \vec \lambda\!\cdot \! \hat v$ reads now
\begin{equation} \Sigma^5=  \Sigma^3 +\zeta\, \Sigma^2+\xi\,\Sigma +\eta\, \mathbf{1}\,. \end{equation} 
 By setting up the linear recursion for five expansion coefficients and solving the matrix exponential function $\exp( i|\vec v\,|M_5)$ one arrives at the following quasi-analytical formula for an $SU(5)$ matrix
 \begin{eqnarray} U = \exp(i \vec \lambda\!\cdot \! \vec v\,) &\!\!\!\!=\!\!\!\!& \sum_{j=1}^5 {\exp(i z_j |\vec v\,|) \over 5z_j^4-3z_j^2-2\zeta z_j-\xi}  \Big\{ (z_j^4-z_j^2-\zeta z_j-\xi)\mathbf{1}\nonumber \\ &&+(z_j^3-z_j-\zeta)\vec \lambda\!\cdot \! \hat v+ (z_j^2-1) (\vec \lambda\!\cdot \! \hat v)^2+ z_j  (\vec \lambda\!\cdot \! \hat v)^3+ (\vec \lambda\!\cdot \! \hat v)^4 \Big\}\,.\end{eqnarray} 
Here $z_1,z_2,z_3,z_4,z_5$ are the five real roots (with zero sum) of the quintic polynomial equation
  \begin{equation} P_5(z) = z^5 -z^3-\zeta z^2-\xi z-\eta = 0\,, \end{equation}
  whose determination with their detailed $(\eta,\zeta,\xi)$-dependence is a formidable task. As the denominator in eq.(25) one recognizes again the derivative $P_5'(z) = 5z^4 -3z^2-2\zeta z-\xi$ evaluated at the root $z_j$.
  
  In perspective one can consider $SU(6)$ with 35 traceless hermitian generators, $\vec \lambda = (\lambda_1, \dots ,\lambda_{35})$ normalized to tr$( \lambda_a \lambda_b)  =2 \delta_{ab}$  (the last one reads $\lambda_{35} ={\rm diag}(1,1,1,1,1,-5)/\sqrt{15}$)  and a 35-component real parameter vector $\vec v = (v_1, \dots ,v_{35})$. The characteristic polynomial that nullifies the rescaled matrix $\Sigma = \vec \lambda \!\cdot \! \hat v$ is of degree six
  \begin{equation}  P_6(z) = z^6-z^4-\zeta z^3-\xi z^2+(\zeta - \chi)z+\eta\,, \end{equation}
   with a new invariant $\chi={\rm tr}\,\Sigma^5/5$. The quasi-analytical formula for $SU(6)$ matrices is analogous to eq.(25) and it involves a sum over the six real roots defined by $P_6(z_j)=0$. The denominator in the formula is $P_6'(z_j) = 6z_j^5-4z_j^3-3\zeta z_j^2-2\xi z_j+\zeta - \chi$ and the coefficients of the expansion with respect to $\bf{1}$ and increasing powers of $ \vec \lambda \!\cdot \! \hat v$ (up to the fifth power) are $z^5-z^3-\zeta z^2-\xi z +\zeta-\chi\,,$ $z^4-z^2-\zeta z-\xi\,,$ $z^3-z-\zeta\,,$ $z^2-1\,,$ $z\,,$ $1\,,$ respectively, each evaluated at the real root $z_j$.
\section{Symplectic group Sp(2)}
The 15-dimensional special unitary group $SU(4)$ contains a particular  10-dimensional subgroup, the so called (compact) symplectic group $Sp(2)$ defined by imposing the condition (of respecting a quaternionic structure) \cite{liegroup}
\begin{equation} U^t J \,U = J\,,  \qquad J =  \left( \begin{array} {cc} \bf{0} & -\bf{1} \\ \bf{1}& \bf{0}\cr \end{array}\!\right),\qquad J^2 =  -\left( \begin{array} {cc} \bf{1} & \bf{0} \\ \bf{0}& \bf{1}\cr \end{array}\!\right), 
\end{equation}  on $U\in SU(4)$, where $^t$ stands for transposition, and $\bf{1}$ and  $\bf{0}$ denote momentarily the $2\!\times\!2$ unit and zero matrix. For the Lie algebra elements $\vec \lambda\!\cdot \!\vec v$ that depend linearly on 15 parameters $v_1,\dots,v_{15}$ this implies the constraint 
\begin{equation}(\vec \lambda\!\cdot \!\vec v\,)^t = J (\vec \lambda\!\cdot \!\vec v\,) J\,, \end{equation}
which as a result eliminates five of the 15 real parameters through the linear relations
\begin{equation} v_9 = v_6\,, \qquad v_{10}=v_7\,,\qquad v_{13}=-v_1\,, \qquad v_{14}=v_2\,,\qquad v_{15}={1 \over \sqrt{2}}\big(v_8-\sqrt{3}v_3\big)\,.\end{equation}
The squared magnitude of the yet 15-component parameter vector $\vec v$  becomes a sum of ten squares
\begin{equation} |\vec v\,|^2 = 2v_1^2+2v_2^2+3\tilde v_3^2+v_4^2+v_5^2+2v_6^2+2v_7^2+\tilde v_8^2+v_{11}^2+v_{12}^2\,,\end{equation}
after introducing the new linear combinations $\tilde v_3=(\sqrt{3}v_3-v_8)/2$ and $\tilde v_8=(v_3+\sqrt{3}v_8)/2$. Moreover, one finds that the trace of the cube vanishes, tr$(\vec \lambda\!\cdot \!\vec v\,)^3=0$, just as a consequence of the five linear relations in eq.(30) or the underlying constraint in eq.(29). Thus one is dealing for the subgroup $Sp(2)$ of $SU(4)$ with the special case $\zeta =0$ and the quartic polynomial $P_4(z) =z^4-z^2+\eta$ becomes biquadratic, and is effectively equivalent to $\tilde P_2(x) = x^2-x+\eta$. The four real roots are then given by 
  \begin{equation}z_{1,3} = \pm \sqrt{x_1}\,, \quad x_1= {1\over 2}\big(1-\sqrt{1-4\eta}\,\big)\geq 0\,,\quad z_{2,4} = \pm \sqrt{x_2}\,, \quad x_2= {1\over 2}\big(1+\sqrt{1-4\eta}\,\big) \geq 0\,,\end{equation}
where the determinant $\eta$ must be  confined to the range $0\leq \eta\leq 1/4$. Since for $Sp(2)$ the roots of $P_4(z)$ occur as pairs of opposite sign, the sum in eq.(15) can be simplified to
\begin{equation}U = \exp(i\vec \lambda\!\cdot \!\vec v\,) = \sum_{j=1}^2 {(x_j-1)\mathbf{1}+(\vec \lambda\!\cdot \!\hat v)^2 \over 2x_j-1}\Big\{\cos( \sqrt{x_j} |\vec v\,|) \mathbf{1} +{i\over \sqrt{x_j}}  \sin(\sqrt{x_j} |\vec v\,|)\,  \vec \lambda\!\cdot \!\hat v\Big\} \,\end{equation} while the trace of such symplectic matrices is always real-valued: tr$\,U = 2 \sum_{j=1}^2  \cos( \sqrt{x_j} |\vec v\,|)$. 

In perspective one can consider the 21-dimensional symplectic group $Sp(3)$ by imposing  on $SU(6)$ matrices the condition $U^t\,J \, U = J$, with $J$ constructed from $3\!\times \!3$ matrices $\bf{1}$ and $\bf{0}$. The condition   $(\vec \lambda\!\cdot \!\vec v\,)^t = J (\vec \lambda\!\cdot \!\vec v\,) J$ for the Lie algebra elements eliminates 14 of the original 35 parameters through linear relations, where the more interesting ones associated to the diagonal generators read\footnote{The remaining 12 relations equate parameters with high and low index: $v_{16}= v_{11}, ~ v_{17}= v_{12}, ~ v_{22}= -v_1, ~v_{23} = v_2,\\ ~v_{25} =v_{13}, ~v_{26} = v_{14}, ~v_{27}= v_{20}, ~v_{28}= v_{21}, ~v_{31} = -v_4, ~ v_{32} = v_5, ~ v_{33} =- v_6, ~ v_{34} = v_7$.}
\begin{equation} v_{24}=  {1\over \sqrt{5}} \big(\sqrt{3} v_{15} - 2 \sqrt{2} v_3\big)\,, \qquad   
 v_{35}= {1\over 2\sqrt{5}} \big(2 \sqrt{2} v_{15} - 5 v_8 - \sqrt{3} v_3\big)\,. \end{equation}
\begin{figure}[ht]\centering
\includegraphics[width=0.45\textwidth]{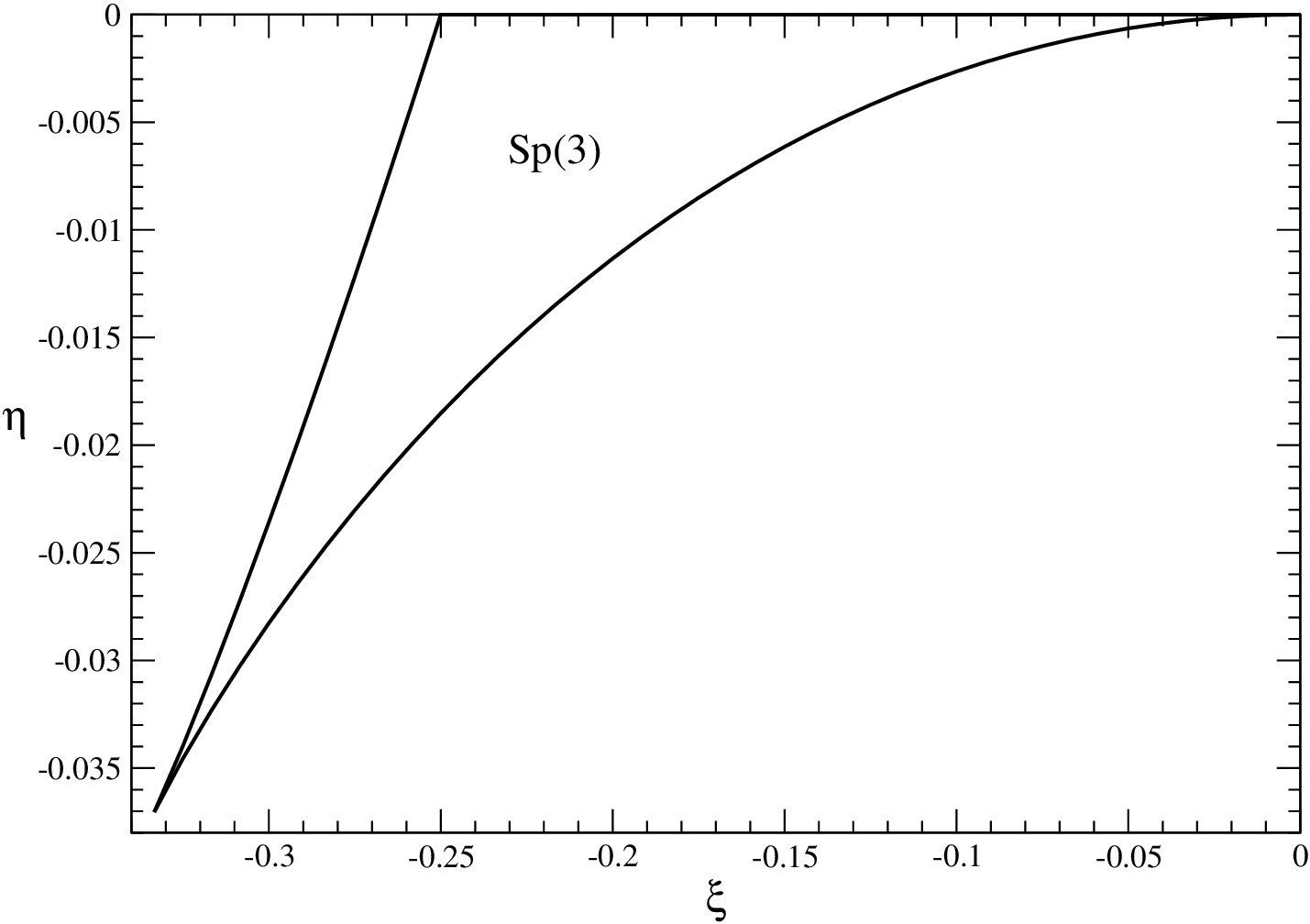}
\caption{The allowed values of the invariants  $\xi$ and $\eta$  for $Sp(3)$ lie between the two curves starting at the point $\xi=-1/3, \eta=-1/27$.} 
\end{figure}
\noindent As a consequence of the imposed condition the traces of odd powers of $\vec \lambda\!\cdot \!\hat v$ vanish, and therefore one is dealing for the subgroup $Sp(3)$ of $SU(6)$ with the special case $\zeta=0,$ $\chi=0$.  The six real roots come as pairs with opposite sign $\pm \sqrt{x_j}$, where $x_j$ are the three positive roots of the cubic polynomial $\tilde P_3(x) = x^3-x^2-\xi x+\eta$. The further analysis based on a behavior of $\tilde P_3(x)$  similar to that shown in Fig.\,2 leads to the following inequalities
\begin{equation} -{1\over 3}\leq \xi \leq 0\,,\quad \eta\leq 0\,, \qquad (2-27\eta+9\xi)^2\leq 4 (1+3\xi)^3\,. \end{equation}
The resulting allowed range of the invariants  $\xi$ and $\eta$ is shown in Fig.\,4 and one recognizes as  the minimal value $\eta_{\rm min} = -1/27$.  Making the substitution  $x={1\over 3}(1+2\sqrt{1+3\xi}\cos\psi)$ one obtains for the three positive roots the expressions
\begin{equation} x_1={1\over 3}(1+2\sqrt{1+3\xi}\cos\psi)\,, \quad x_2={1\over 3}\Big[1-2\sqrt{1+3\xi}\sin\Big(\psi+{\pi \over 6}\Big)\Big] \,, \quad x_3={1\over 3}\Big[1+2\sqrt{1+3\xi}\sin\Big(\psi-{\pi \over 6}\Big)\Big]\,, \end{equation}
with the angle $\psi\in[0,\pi/3]$ given by
\begin{equation}\psi = {1\over 3}\arccos{2-27\eta+9\xi \over 2(1+3\xi)^{3/2}}\,.\end{equation}
In the end the semi-analytical formula for $Sp(3)$ matrices reads
\begin{eqnarray}U = \exp(i\vec \lambda\!\cdot \!\vec v\,) &\!\!\!\!\!=\!\!\!\!&\sum_{j=1}^3 {1 \over 3x_j^2-2x_j-\xi}\Big\{\cos( \sqrt{x_j} |\vec v\,|) \mathbf{1} +{i\over \sqrt{x_j}}  \sin(\sqrt{x_j} |\vec v\,|)\,\vec \lambda\!\cdot \!\hat v\Big\}\nonumber \\ && \qquad  \times \Big\{ (x_j^2-x_j-\xi) \mathbf{1} +(x_j-1)(\vec \lambda\!\cdot \!\hat v)^2+ (\vec \lambda\!\cdot \!\hat v)^4\Big\} \,,\end{eqnarray} 
where the trace tr$\,U = 2\sum_{j=1}^3 \cos( \sqrt{x_j} |\vec v\,|) $ is again real-valued.
 
In passing one reminds that in low dimensions the spin groups $Spin(n)$, defined as the two-sheeted simply-connected coverings of the special orthogonal goups $SO(n)$, obey the following isomorphisms
 \begin{equation}Spin(3)= SU(2)\,, \quad Spin(4) = SU(2) \!\times \!SU(2)\,, \quad Spin(5) =Sp(2)\,, \quad Spin(6) = SU(4)\,,\end{equation}
together with $Sp(1)= SU(2)$. For all these compact Lie groups the analytical evaluation of the matrix exponential function has been studied in this work. Actually, the obtained formula can be  evaluated most straightforwardly for the symplectic group $Sp(2)$, whereas in the other cases one has to make a (somewhat) cumbersome detour via the three real roots of a cubic polynomial equation. 
\subsection*{Acknowledgement} This work  has been 
supported in part by DFG (Project-ID 196253076 - TRR 110) and NSFC.


\begin{thebibliography}{99}
\bibitem{ulfbook} Ulf-G. Mei{\ss}ner and A. Rusetsky, {\it Effective Field Theories}, Cambridge University Press (2022).\vs
\bibitem{koecher} M. K\"ocher, {\it Linear Algebra and Analytical Geometry}, Springer Verlag (2003), paragraphs 3.4.6 and 8.3.9.\vs
\bibitem{algebra} B.L. van der Waerden, {\it Algebra I}, Springer Verlag (1971), paragraph 64.\vs\bibitem{liegroup} Th. Br\"ocker and T. tom Dieck, {\it Representations of Compact Lie Groups}, Springer Verlag (1985), chapter 1.\vs
\end{thebibliography}
\end{document}